\documentclass[a4paper,fleqn,usenatbib]{mnras}

\usepackage[T1]{fontenc}
\usepackage{ae,aecompl}

\usepackage{graphicx}
\usepackage{amsmath}
\usepackage{amssymb}

\title[Relativistic jets in the intracluster medium]
      {Relativistic hydrodynamic jets in the intracluster medium}

\author[E. Choi]
       {Eunwoo Choi$^{1}$\thanks
       {Present address: Exhibition Research Division,
        Daegu National Science Museum, Daegu 43023, Korea} \\
  $^{1}$Department of Astronomy and Atmospheric Sciences,
        Kyungpook National University, Daegu 41566, Korea}

\date{Accepted 2017 ***. Received 2017 ***; in original form 2017 ***}

\pubyear{2017}

\begin{document}

\label{firstpage}
\pagerange{\pageref{firstpage}--\pageref{lastpage}}

\maketitle

\begin{abstract}
We have performed the first three-dimensional relativistic hydrodynamic
simulations of extragalactic jets of pure leptonic and baryonic plasma
compositions propagating into a hydrostatic intracluster medium
environment.
The numerical simulations use a general equation of state for a
multi-component relativistic gas which closely reproduces the Synge
equation of state for a relativistic perfect gas.
We find that morphological and dynamical differences between leptonic
and baryonic jets are much less evident than those between hot and cold
jets.
In all these models, the jets first propagate with essentially constant
velocities within the core radius of the intracluster medium and then
accelerate progressively so as to increase the jet advance velocity by
a factor of between 1.2 and 1.6 at the end of simulations, depending upon
the models.
The temporal evolution of the average cavity pressure is not consistent
with that expected by the extended theoretical model even if the average
cavity pressure decreases as a function of time with a power-law.
Our simulations produce synthetic radio images that are dominated by
bright hot spots and appear similar to observations of the extended
radio galaxies with collimated radio jets.
These bright radio lobes would be visible as dark regions in X-ray images and are
morphologically similar to observed X-ray cavities in the intracluster
medium.
This supports the expectation that the bow shock surrounding the head of
the jet is important mechanism for producing X-ray cavities in the
intracluster medium.
Although there are quantitative differences among the models the total
radio and X-ray intensity curves show qualitatively similar trends in
all of them.
\end{abstract}


\section{Introduction}

A substantial number of observational and theoretical studies show that
powerful extragalactic radio sources have highly relativistic and
collimated jets that emerge from a central active galactic nucleus (AGN).
These relativistic jets transport away bulk kinetic energy from the AGN
to the lobes; this kinetic energy is dissipated by shocks within the beam
and at the terminal shocks where electrons are accelerated \citep{bla74}.
In the current paradigm, these relativistic jets could be made of a
baryonic plasma containing relativistic electrons and thermal protons,
a leptonic plasma of relativistic electrons and positrons, or possibly
other mixtures of relativistic electrons and other species.
Since the plasma compositions of extragalactic jets are directly related
to their formation mechanism, stability, and energy transport, the
composition of jets has been recognized as important for a long time,
but it is still poorly known \citep[e.g.,][]{dey06}.

There is strong observational evidence that the intracluster medium (ICM)
interacts with the radio jets
\citep[e.g.,][]{fab00,fab06,wis07,for07,bla09,ran11,san16}.
This conclusion has been driven by X-ray observations, which have shown
clearly that a powerful radio source affects the ICM significantly in
the vicinity of the source.
In particular, the propagating relativistic jets inflate large underdense
cavities which displace the local intracluster hot gas and hence appear
as dark holes in X-ray observations.
The observed radio lobes are coincident with evacuated cavities that are
bounded by shells of enhanced emission in the intracluster medium.
Occasionally in the hot ICM, apparent relic extended radio structures
from the central AGN are observed with no, or very faint, apparent radio
connections.
These various phenomena imply that outflows from AGN can inflate X-ray
cavities or radio bubbles in the ICM, which then may dynamically evolve
on their own \citep[e.g.,][]{ran11}.

An substantial effort toward modeling the interaction between radio jets
and the ICM has been made using hydrodynamic simulations
\citep[e.g.,][]{rey01,rey02,bas03,omm04,ver06,hei06,cat07,mor10,gas11,har13}.
These non-relativistic hydrodynamic simulations give us a reasonable
outline of how the radio sources impact the ICM and of how they evolve.
\citet{rey01,rey02} simulated, in two dimensions, AGN radio jets in a
hydrostatic, spherically symmetry galaxy cluster model.
\citet{mor10} performed three-dimensional simulations of AGN jets in a
cosmologically evolved, realistic galaxy cluster.
They found that large-scale motions of cluster gas disrupt the AGN jets,
which leads to the appearance of multiple disconnected X-ray bubbles
from a AGN with a constant luminosity.
Although the basic physical processes have been established by these
numerical studies, the details of the evolution of a relativistic jet in
a cluster atmosphere needs to be fully understood.
In the context of relativistic hydrodynamic simulations, several
numerical simulations of the interaction of the relativistic jets with
the surrounding material have been reported
\citep{kom98,sch02,mel08,ros08,per11,per14}.
These simulations use a more general equation of state to account for
transitions from non-relativistic to relativistic regimes.
\citet{sch02} used two-dimensional relativistic hydrodynamic simulations
to investigate the influence of the composition of relativistic jets on
their long-term evolution in a homogeneous external medium.
\citet{mel08} studied the relativistic jet deceleration through density
discontinuities using two-dimensional hydrodynamic simulations.
More recently, \citet{per14} performed two-dimensional hydrodynamic
simulations of relativistic jets propagating into a cluster environment
to attempt to understand ICM heating.

Since the relativistic flows change their state, a realistic equation of
state must be adopted to handle both nonrelativistic and relativistic
states.
The exact form of an equation of state relating thermodynamic quantities
of specific enthalpy and temperature is completely described in terms of
modified Bessel function \citep{syn57}.
The Synge equation of state for the relativistic perfect gas, however,
is not efficient and thus unsuitable from the computational point of
view since it involves the computation of Bessel functions, which entail
a significant computational cost, especially in three-dimensional
high-resolution simulations.
To avoid the direct use of the Synge equation of state, \citet{mig07}
used, in their numerical relativistic hydrodynamic code, a simple and
linear equation of state that closely approximates the Synge equation of
state for a single-component relativistic gas.
In the context of the relativistic hydrodynamic code, \citet{cho10}
proposed a general equation of state for a multi-component relativistic
gas which is consistent with the Synge equation of state for a
relativistic perfect gas and is efficient and suitable for numerical
relativistic hydrodynamics.
This proposed general equation of state has an analytic expression and
closely reproduces the Synge equation of state for a multi-component
relativistic gas in all regimes.

In this paper we present the first three-dimensional relativistic
hydrodynamic simulations of extragalactic jets with different plasma
compositions in an inhomogeneous ICM environment, using a general
equation of state.
We focus not only on how the relativistic jet interaction with an ICM
affects the morphology, dynamics, and emission of the jets and the ICM
but also how the different plasma compositions of the jets influence
their evolutions.
The paper is organized as follows.
In Section 2 we describe the details of the numerical simulations of our
work, and in Section 3 we discuss and interpret the simulation results.
A discussion and conclusion are presented in Section 4.

\section{Numerical simulations}

\subsection{Relativistic hydrodynamic equations}

The equations of special relativistic hydrodynamics with weak gravity
\citep{hwa16} can be written as conservation laws for mass, momentum and
energy as
\begin{equation}
\frac{\partial D}{\partial t}+\bmath{\nabla}\cdot\left(D\mathbfit{v}\right) = 0,
\end{equation}
\begin{equation}
\frac{\partial\mathbfit{M}}{\partial t}+\bmath{\nabla}\cdot\left(\mathbfit{Mv}+p\mathbf{I}\right)
= \left(E+p\right)\bmath{\nabla}\phi,
\end{equation}
\begin{equation}
\frac{\partial E}{\partial t}+\bmath{\nabla}\cdot\left[\left(E+p\right)\mathbfit{v}\right]
= \left(E+p\right)\mathbfit{v}\cdot\bmath{\nabla}\phi.
\end{equation}
Here the conservative variables $D$, $\mathbfit{M}$, and $E$ represent
respectively the mass density, the momentum density, and the energy
density in the reference frame, the primitive variables $\rho$,
$\mathbfit{v}$, and $p$ denote respectively the rest mass density,
velocity, and pressure in the local rest frame, $\mathbf{I}$ is the
$3\times3$ unit tensor, and $\phi$ is the gravitational potential.
The speed of light is set to unity ($c=1$) throughout this work.

The conservative variables measured in the reference frame are related
to the primitive variables in the local rest frame according to the
transformations
\begin{equation}
D = \Gamma\rho,
\end{equation}
\begin{equation}
\mathbfit{M} = \Gamma^2\rho h\mathbfit{v},
\end{equation}
\begin{equation}
E = \Gamma^2\rho h-p,
\end{equation}
where $\Gamma = 1/\sqrt{1-v^2}$ is the Lorentz factor and $h$ is the
specific enthalpy.
The transformation is nonlinearly coupled and reduces to a single
nonlinear equation.

The system of conservation equations is closed by an equation of state
that can be expressed for the specific enthalpy as a function of the
rest mass density and the pressure $h = h(\rho,p)$.
As mentioned earlier, the exact form of equation of state for a
relativistic perfect gas was given by \citet{syn57} and the Synge
equation of state is entirely described in terms of the modified Bessel
function.
\citet{mat71} introduced a linear equation of state which closely
reproduces the Synge equation of state for a single-component
relativistic gas.
For a multi-component relativistic gas composed of electrons, positrons,
and protons, \citet{cho10} studied the linear approximation of the Synge
equation of state and proposed a new general equation of state that uses
an analytical expression and is suitable for numerical computations.
We adopt that proposed equation of state which takes the form
\begin{displaymath}
h = \frac{5}{2}\frac{1}{\xi}
+\left(2-\chi\right)\left[\frac{9}{16}\frac{1}{\xi^2}+\frac{1}{\left(2-\chi+\chi\mu\right)^2}\right]^{1/2}
\end{displaymath}
\begin{equation}
~~~~~+\chi\left[\frac{9}{16}\frac{1}{\xi^2}+\frac{\mu^2}{\left(2-\chi+\chi\mu\right)^2}\right]^{1/2},
\end{equation}
where $\xi = \rho/p$ is a measure of inverse temperature,
$\chi = n_{p^+}/n_{e^-}$ is the relative fraction of proton and electron
number densities, and $\mu = m_p/m_e$ is the mass ratio of proton to
electron.
We note that $\chi = 0$ represents an electron-positron gas while
$\chi = 1$ indicates an electron-proton gas.

The explicit form of the sound speed depends on the equation of state.
The sound speed is written as
\begin{equation}
c_s^2 = \frac{\gamma_r p}{\rho h},
\end{equation}
and the relativistic adiabatic index is given by
\begin{equation}
\gamma_r = \frac{h^\prime\xi^2}{h^\prime\xi^2+1},
\end{equation}
where $h^\prime = dh/d\xi$.
The relativistic adiabatic index is constant ($\gamma_r = 4/3$ or $5/3$)
if the gas remains ultrarelativistic ($\xi\ll1$) or subrelativistic
($\xi\gg1$).
For the intermediate regime the quantity $\gamma_r$ varies between the
two limiting cases.

The simulation results are given in arbitrary units.
We can relate the dimensionless quantities to the physical quantities by
fixing the relevant parameters of the ICM to values representative of a
typical galaxy cluster.
In this work we can use this scaling when converting our simulation
results into physical quantities that can be compared directly with real
systems.

\subsection{Intracluster medium}

The ICM can be reasonably treated as a spherical halo consisting of a
hot gas and stationary dark matter and its dynamics is followed in the
gravitational potential of background dark matter.
The hot gas is assumed to be in hydrostatic equilibrium and energy
dissipation due to radiative cooling is excluded.
The radial density profile of the dark matter halo can be taken to be
described by the NFW model \citep{nav97}.
Several other choices such as a power-law distribution are also
available for modeling the mass distribution of the ICM, including dark
matter.

The gas density distribution of the ICM is usually considered to be
given by a $\beta$-model profile
\begin{equation}
\rho(r) = \frac{\rho_0}{\left[1+\left(r/r_0\right)^2\right]^{3\beta/2}},
\end{equation}
where $r$ is the spherical radius from the center of the ICM, $\rho_0$
is the mass density at the center of the ICM, $r_0$ is its core radius,
and the $\beta$ parameter describes the rate of decrease of density at
large distances.
We set $\rho_0 = 1$, $r_0 = 1$, and $\beta = 0.5$ in this work.
The $\beta$-model is a simple analytic fit to the mass profile of the
ICM, assuming it is spherically symmetric and isothermal, rather than a
fully realistic realization of the density structure in a particular
cluster.

The hot gas of the ICM is in spherical hydrostatic equilibrium within
the gravitational potential of the static dark matter halo
\begin{equation}
\bmath{\nabla}p = \rho\bmath{\nabla}\phi.
\end{equation}
The dynamical equilibrium of the gas is attained by introducing an
external force which compensates initial pressure gradients in all the
directions and keeps the gas stationary everywhere.

In the nonrelativistic regime ($\xi\gg1$) applicable to the ICM, the
thermodynamic variables become $h\rightarrow1$,
$c_s\rightarrow1/\sqrt{\xi}$, and $\gamma_r\rightarrow5/3$, so these
variables are assumed to be constant in the ICM.
Following \citet{rey02}, these assumptions give us the gravitational
potential as
\begin{equation}
\phi(r) = \frac{h_a c_{s,a}^2}{\gamma_{r,a}}\ln\left[\rho(r)\right],
\end{equation}
where the subscript $a$ stands for the ambient gas and the thermodynamic
variables are set to $h_a  = 1$, $c_{s,a} = 0.01$, and $\gamma_{r,a} = 5/3$.
Since the gravitational potential is determined by the mass distribution
of the background dark matter, the self-gravity of the ICM can be taken
to be negligible.

\begin{figure}
\includegraphics[width=\columnwidth]{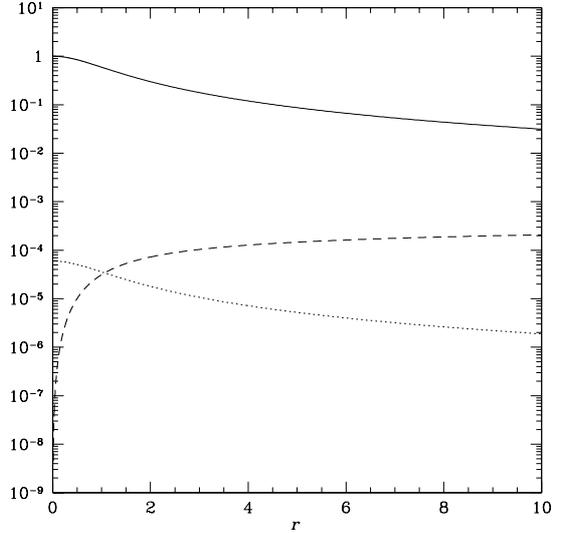}
\caption{The initial profiles of gas density (solid curve), gravitational potential (dashed curve),
and gas pressure (dotted curve) along the radial direction (with $r_0 = 1$) in the $\beta$-model of the
ICM.}
\label{fig1}
\end{figure}

The initial distributions of gas density, gravitational potential, and
gas pressure along the radial direction are shown in Figure \ref{fig1}.
The hydrostatic equilibrium is initially maintained in all the directions
under the gravity and this keeps the structure of the ICM stationary.
Since we set $c = 1$, $\rho_0 = 1$, and $r_0 = 1$ in our models, the
units of all physical quantities are arbitrary and can be easily
converted to any specific physical units.

\subsection{Numerical models}

To carry out the numerical simulations of the relativistic jets in the
ICM, we used the multidimensional relativistic hydrodynamic code
described in \citet{cho10}.
The numerical code is based on the HLL scheme \citep{har83} with the
estimates of the wave speeds from \citet{ein88} and is fully explicit
and second-order accurate in space and time.
It incorporates the analytical formulation of the general equation of
state for a multi-component relativistic gas, so that it is efficient
and suitable for numerical relativistic hydrodynamics and gives accurate
thermodynamic results for relativistic astrophysical flows.
The code was extensively tested and its ability was demonstrated in a
variety of relativistic test problems including highly relativistic
shock tubes with tangential velocities.

The jet models considered here include a hot leptonic jet, a hot baryonic
jet, a cold leptonic jet, and a cold baryonic jet defined according to
the beam internal energy and the plasma composition of the jet.
The models have beam density $\rho_b = 10^{-4}$ for the hot models and
$10^{-3}$ for the cold models, beam velocity $v_b = 0.9$, and beam
pressure $p_b>p_0$ for the hot models and $\leq p_0$ for the cold models,
where $p_0 = 10^{-4}$ is the pressure at the center of the ICM.
The model parameters correspond to a density contrast
$\eta = \rho_b/\rho_0 = 10^{-4}$ for the hot models and $10^{-3}$ for
the cold models, beam Lorentz factor $\Gamma_b = 2.29$, and pressure
ratio $K = p_b/p_0>1$ for the hot models and $\leq1$ for the cold models.
This gives the beam Mach number $M_b = v_b/c_{s,b} = 1.57$ and $1.58$
for the hot models, where $c_{s,b}$ is the beam sound speed and
$M_b = 2.55$ and $2.97$ for the cold models.
We set $\chi = 0$ (electron-positron gas) for the leptonic models and
$\chi = 1$ (electron-proton gas) for the baryonic models.
Since our code holds in the limit where the composition of the plasma is
fixed in space and time, the composition $\chi$ is evolved by a constant
value in our simulations.
In reality, however, the plasma composition may change due to fluid
mixing or the creation or annihilation of electron-positron pairs.
Nonetheless, this assumption of constant $\chi$ is a useful first order
approximation in these simulations.
The computational domain is a three-dimensional box spanning a size of
$x = [0,2]$, $y = [0,2]$, and $z = [0,4]$ in Cartesian geometry with an
uniform grid of $256\times256\times512$ cells.
The jet has an initial beam radius $R_b = 1/16$ (corresponding to $8$
cells), and is launched from an injection region at the origin where an
inflow boundary condition is imposed, and propagates through the ambient
medium along the $z$-direction.
Outflow boundary conditions are used at all boundaries except along the
symmetry axis where reflecting boundary conditions are taken for the jet
axis.

The kinetic luminosity of the beam obtained by integrating the energy flux
over the beam cross section \citep{sch02} is given by
\begin{equation}
L_b = \left(\Gamma_b h_b-1\right)\Gamma_b\rho_b v_b\pi R_b^2.
\end{equation}
We have computed the kinetic luminosity of the beam by fixing the
parameters of the background medium.
We set $\rho_0 = 0.01$\,cm$^{-3}$ and $r_0 = 10$\,kpc which are general
values representative of galaxy clusters.
The kinetic luminosity of the beam is then
$L_b = 2.0\times10^{46}$\,erg\,s$^{-1}$ in all the models, which is
a typical value for a powerful AGN jet in a cluster.

The physical parameters of each model are summarized in Table \ref{tab1}.
In our simulations we considered four models of relativistic jets in the
ICM as well as one model of the ICM with no injected jet to check that
the background medium remains in hydrostatic equilibrium over the
simulation time.

\begin{table}
\caption{Summary of physical parameters for simulation models.}
\label{tab1}
\begin{tabular}{lccccc}
\hline
 Model          & IC       & HL        & HB        & CL        & CB        \\
\hline
 $\eta$         & --       & $10^{-4}$ & $10^{-4}$ & $10^{-3}$ & $10^{-3}$ \\
 $\Gamma_b$     & --       & 2.29      & 2.29      & 2.29      & 2.29      \\
 $K$            & --       & 2.13      & 2.09      & 1.00      & 0.80      \\
 $M_b$          & --       & 1.57      & 1.58      & 2.55      & 2.97      \\
 $\gamma_{r,b}$ & --       & 1.34      & 1.35      & 1.58      & 1.43      \\
 EOS            & $e^-p^+$ & $e^-e^+$  & $e^-p^+$  & $e^-e^+$  & $e^-p^+$  \\
 $\chi$         & 1.0      & 0.0       & 1.0       & 0.0       & 1.0       \\
 $R_b$          & --       & 1/16      & 1/16      & 1/16      & 1/16      \\
 $t$            & 10       & 45        & 50        & 35        & 30        \\
 $L_b$ [$10^{46}$\,erg\,s$^{-1}$]
                & --       & 2.0       & 2.0       & 2.0       & 2.0       \\
\hline
\end{tabular}

\medskip
{Note - The model designation indicates a hot (H) or cold (C) jet
followed by a leptonic (L) or baryonic (B) plasma; the model of the
ICM only (no jet injected) is designated by IC.
The rows give from top to bottom the density contrast $\eta$, the
beam Lorentz factor $\Gamma_b$, the pressure ratio $K$, the beam Mach
number $M_b$, the relativistic adiabatic index of the beam
$\gamma_{r,b}$, the type of the equation of state EOS, the relative
fraction of proton and electron number densities $\chi$, the beam radius
$R_b$, the simulation time $t$, and the kinetic luminosity of the beam
$L_b$ in units of $10^{46}$\,erg\,s$^{-1}$.}
\end{table}

\begin{figure}
\includegraphics[width=\columnwidth]{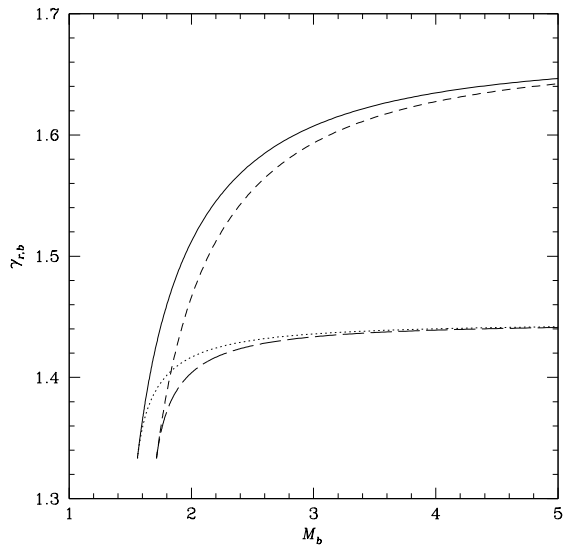}
\caption{The relativistic adiabatic index of beam as a function of the
beam Mach number for the leptonic and baryonic models.
Short dashed and long dashed lines correspond to the leptonic and
baryonic models with $v_b = 0.9$, respectively, while solid and dotted
lines show the leptonic and baryonic models with $v_b = 0.99$ for
comparison.}
\label{fig2}
\end{figure}

Figure \ref{fig2} shows the relativistic adiabatic index of the beam,
$\gamma_{r,b}$, as a function of the beam Mach number, $M_b$, for the
leptonic and baryonic models with different beam velocities.
Discrepancies between the leptonic and baryonic models increase at high
beam Mach numbers and reduce at low beam Mach numbers, where the value
of $\gamma_{r,b}$ asymptotically approaches $4/3$.
In our models the hot leptonic and baryonic models correspond to
$\gamma_{r,b} = 1.34$ and $1.35$ and the cold leptonic and baryonic
models correspond to $\gamma_{r,b} = 1.58$ and $1.43$, respectively.

In order to track the evolution of the jet material we employed a
passive mass fraction tracer $f$ which is evolved by means of an
additional conservation equation
$\partial(Df)/\partial t+\bmath{\nabla}\cdot(Df\mathbfit{v}) = 0$.
The mass fraction tracer is assigned to $f = 1$ for material originating
inside the beam injection region and to $f = 0$ for ambient material
elsewhere.
We defined the beam density $\rho_b = \rho f_b$ as the region where
$f_b\geq0.9$ and the cavity density $\rho_c = \rho f_c$ as the region
where $f_c>0$.
When the numerical code integrates the conservation equation
$\partial\mathbfit{q}/\partial t+\bmath{\nabla}\cdot\mathbf{F} = \mathbfit{S}$,
where $\mathbfit{q}$ is the state vector for the conservative variables,
$\mathbf{F}$ is the flux tensor, and $\mathbfit{S}$ is the source vector
introduced by gravity, the contribution of the source terms is included
through a separate advection step in the code.
That is, the state vector $\mathbfit{q}$ is first updated by solving
$\partial\mathbfit{q}/\partial t+\bmath{\nabla}\cdot\mathbf{F} = 0$, via
operator splitting and then the source vector $\mathbfit{S}$ is added by
solving $\partial\mathbfit{q}/\partial t = \mathbfit{S}$, during the advection
step.

\section{Numerical results}

We have run a total of five model simulations whose global parameters
are summarized in Table \ref{tab1}.
The first one, model IC, is a purely hydrostatic simulation, where we
have verified that the density distribution of the ICM remains constant
for the duration of the run.
As the first simulation is just a check on the accuracy of our
hydrostatic equilibrium, we will only discuss the other four simulations.
The four simulations are of the evolution of relativistic jets with
different compositions propagating through a decreasing density
atmosphere of the ICM.
The numerical resolution adopted in all the three-dimensional simulations
does not allow for very detailed study of the complex fluctuations
generated by the outflows.
However, this resolution is comparable enough to resolve the jet dynamics
and cavity evolution and appropriate for investigating the global
properties of the outflows such as the averaged pressure profiles and
total radio and X-ray intensity curves, which we discuss in the following
subsections.

\subsection{Dynamic evolutions}

\begin{figure*}
\includegraphics[width=\columnwidth]{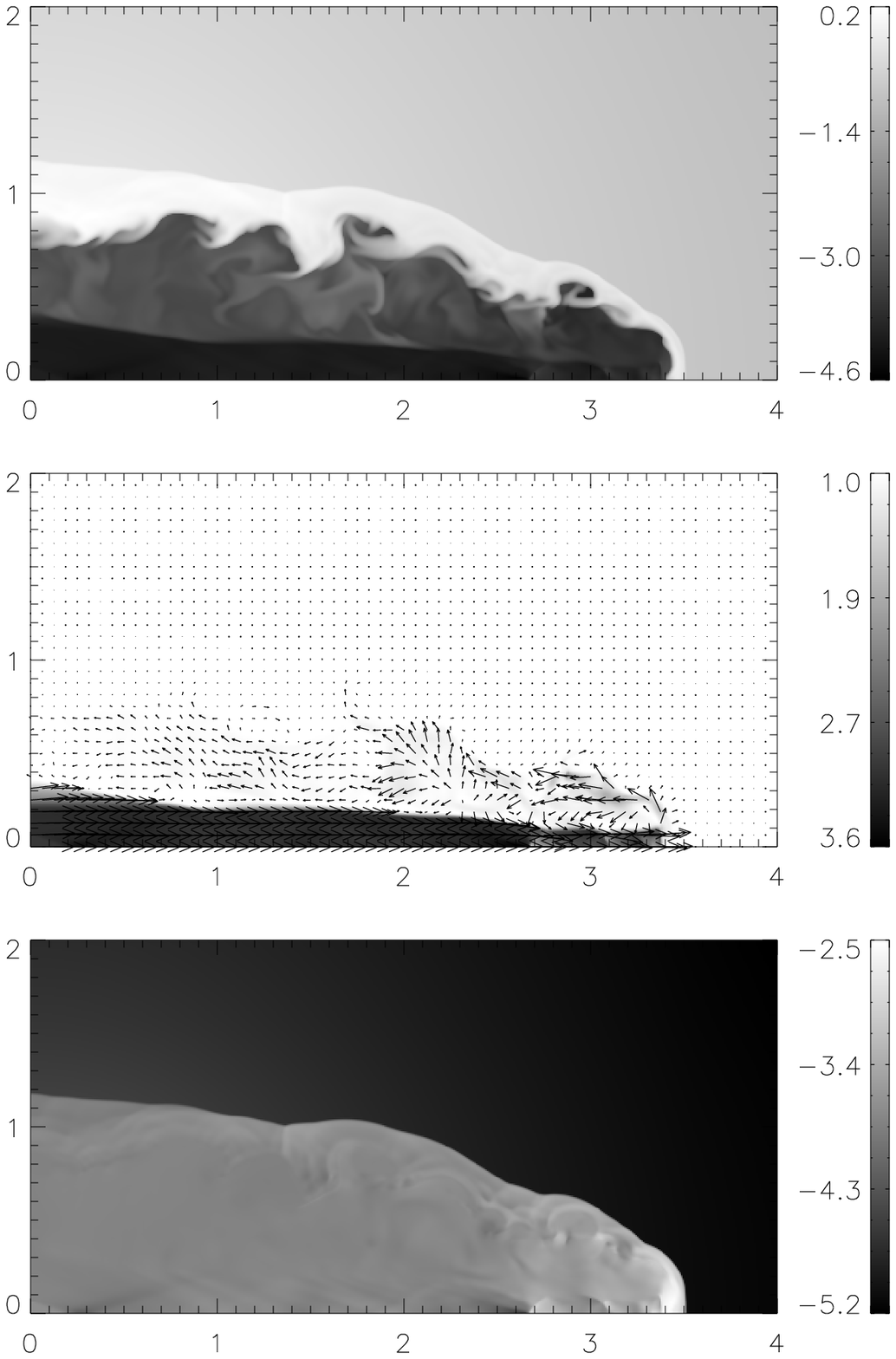}
\includegraphics[width=\columnwidth]{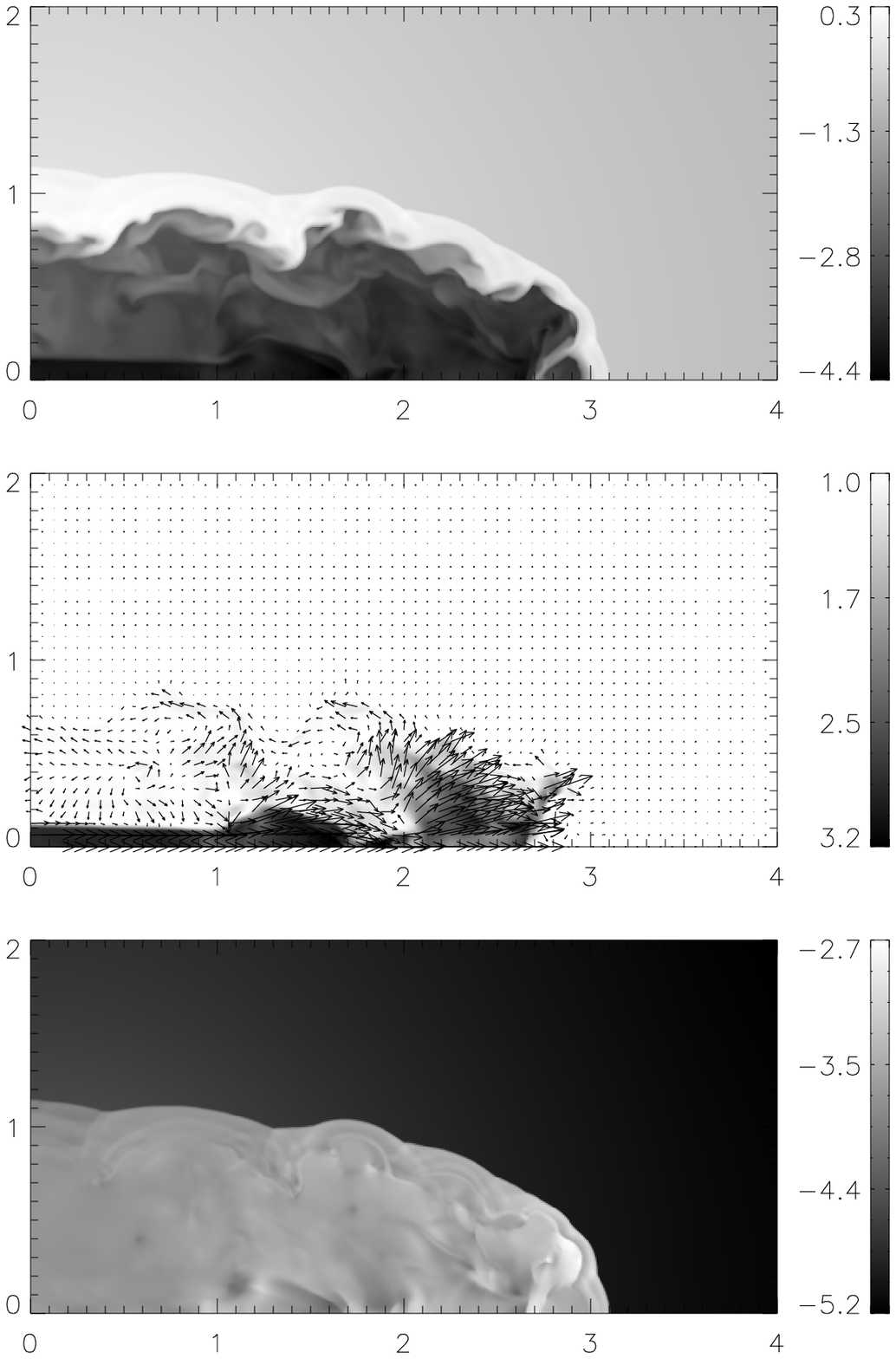}
\caption{Images of the rest mass density, the Lorentz factor with
velocity field, and the pressure (top to bottom) for the hot leptonic
model HL (left) and the hot baryonic model HB (right).
The images of the rest mass density and the pressure are shown in
logarithmic scales and the velocity field overlaid on the Lorentz factor
is normalized with the longest velocity vector.
All the images are shown on the plane $x = 0$ in the three-dimensional
Cartesian geometry at time $t = 45$.}
\label{fig3}
\end{figure*}

The images in Figure \ref{fig3} display the logarithm of the rest mass
density, the Lorentz factor overlaid with the velocity field, and the
logarithm of the pressure on the plane $x = 0$ in the three-dimensional
computation domain at time $t = 45$ for the hot leptonic jet of model HL
and the hot baryonic jet of model HB.
For each model, a bow shock that separates the shocked jet material from
the shocked ambient medium is driven into the ambient medium.
The beam itself is terminated by a Mach disk where much of the beam
kinetic energy is converted into internal energy.
The shocked jet material flows backward along the working surface into a
cocoon, resulting in the development of turbulent vortices in the cocoon.
The interaction of these vortices with the beam form oblique internal
shocks inside the beam close to the terminal Mach disk, which causes the
deceleration of the jet.
These overall morphological features of hot leptonic and baryonic jets
are similar to those of the hot relativistic jets with a constant
adiabatic index studied by \citet{cho07}.
The models HL and HB, however, show some features that differentiate
them in morphological and dynamical aspects.
The hot leptonic jet propagates relatively faster, with the average
velocity of jet head $v_h\sim0.08$ while the hot baryonic jet has
$v_h\sim0.07$.
The hot leptonic jet model features a relatively narrower conical
shape for the bow shock, while the hot baryonic jet model has a bow shock
with a broader conical shape.
The shapes of the cocoons and the internal structures of the beams are
similar between the models HL and HB, even though they have different
plasma compositions.

\begin{figure*}
\includegraphics[width=\columnwidth]{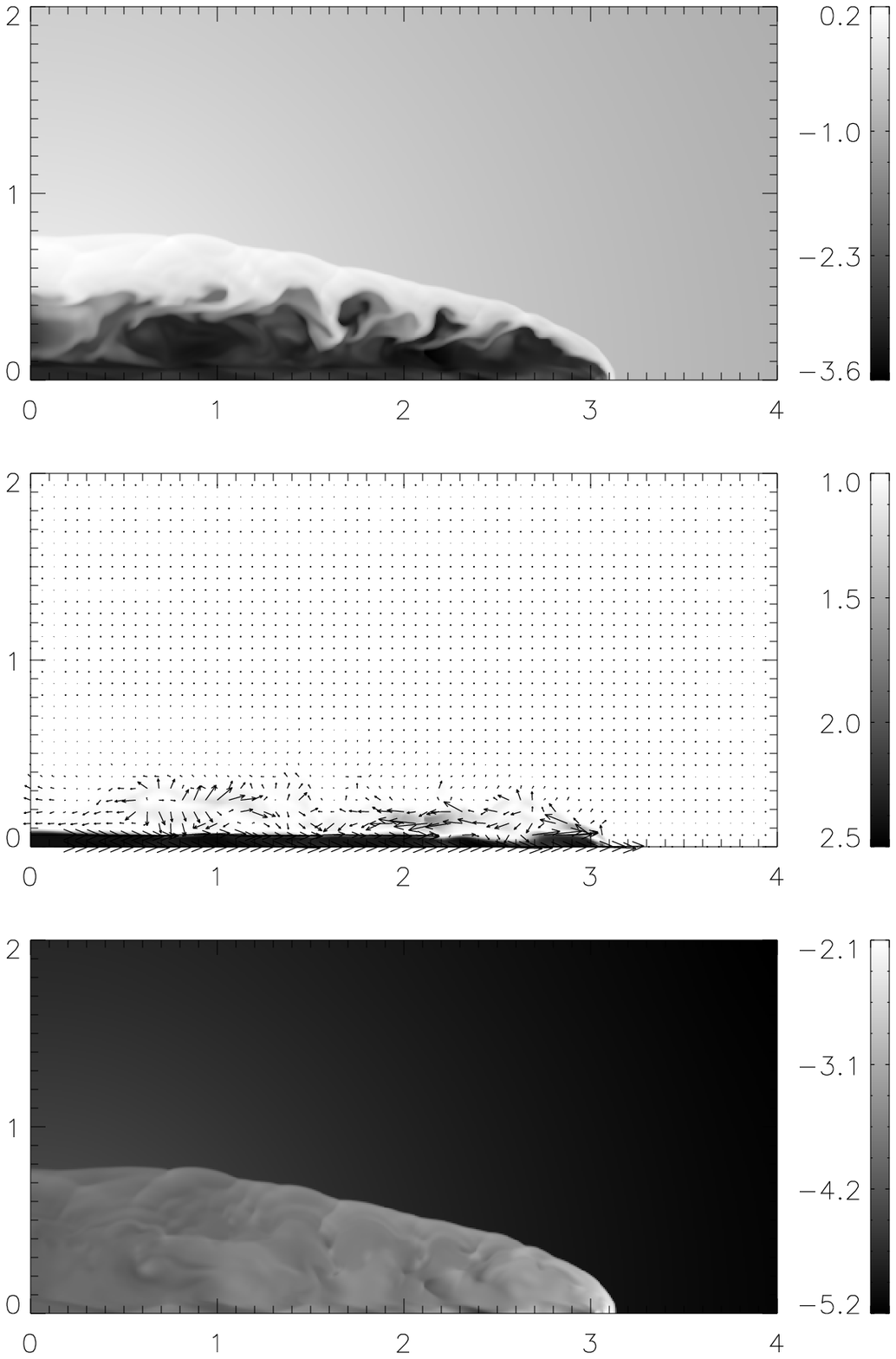}
\includegraphics[width=\columnwidth]{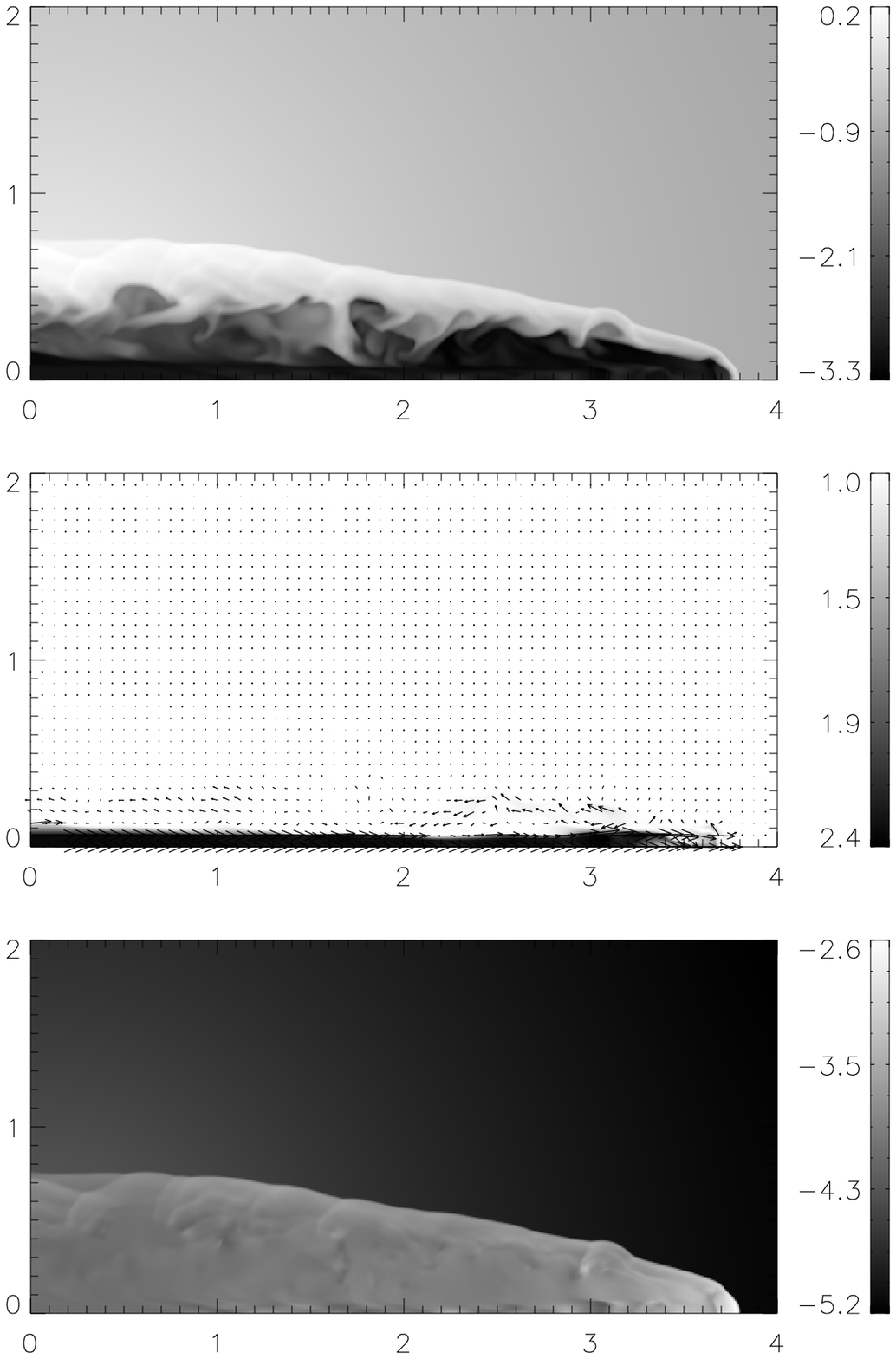}
\caption{Images of the rest mass density, the Lorentz factor with
velocity field, and the pressure (top to bottom) for the cold leptonic
model CL (left) and the cold baryonic model CB (right).
The images of the rest mass density and the pressure are shown in
logarithmic scales and the velocity field overlaid on the Lorentz factor
is normalized with the longest velocity vector.
All the images are shown on the plane $x = 0$ in the three-dimensional
Cartesian geometry at time $t = 30$.}
\label{fig4}
\end{figure*}

The images in Figure \ref{fig4} show the logarithm of the rest mass
density, the Lorentz factor with velocity field overlaid, and the
logarithm of the pressure on the plane $x = 0$ in the three-dimensional
computation domain at time $t = 30$ for the cold leptonic jet of model
CL and the cold baryonic jet of model CB.
The models CL and CB show specific differences in morphology and
dynamics, which are dissimilar from the differences between the models HL
and HB.
The cold leptonic jet of model CL advances at the relatively slower
velocity $v_h\sim0.10$ than does the cold baryonic jet of model CB with
$v_h\sim0.13$.
The cold leptonic jet generates a slightly broader conical shape to a
bow shock, and the cold baryonic jet makes a bow shock with a narrower
conical shape.
Although the morphological and dynamical differences in the jet head are
distinguishable, the shapes of the cocoons and the internal structures
of the beams are similar between the models CL and CB with different
plasma compositions.
In spite of their very different plasma compositions, the differences in
morphology and dynamics are not significant between the leptonic and
baryonic jets.
In this investigation, the morphological and dynamical differences
between the hot and cold jets are much more evident than those between
the leptonic and baryonic jets.
The cold jets propagate at faster speeds and produce narrower bow shocks
and thin cocoons; on the contrary, the hot jets are characterized by slower
advance velocities and broader bow shocks and thicker cocoons.
The terminal Mach disk stands off farther behind the bow shock in the
hot jet models while it is located quite close to the bow shock in the
cold jet models.
There are some differences in the evolution of the pressure between these
simulations and those done by \citet{sch02} and \citet{per14}.
The discrepancies are presumably caused by the use of different simulation
parameters for the jets and ICM models and also by the adoption of different equation of
state, but it is not possible to establish the dominant effect without conducting a
substantial number of additional simulations.

We introduced several quantities to characterize some dynamical and
morphological aspects of the different models.
The axial length of the cavity $l_c$ is the head position of the jet,
the $z$-coordinate of the leading bow shock $z_c$, in which
$\triangle v_z<0$ and $\vert\triangle p\vert/p>1$.
The radial size of the cavity $R_c = \sqrt{V_c/(\pi l_c)}$ is the radial
dimension of a cylinder that has the same volume $V_c$ and length $l_c$
as the cavity.
The aspect ratio of the cavity is then defined by $A_c = l_c/R_c$.
A simple theoretical model for the evolution of the cocoon or cavity of
AGN jets was described in \citet{beg89} and it was extended by
\citet{sch02} and \citet{per11}.
In this extended model, if the ambient density follows the power law
$\rho\propto r^{-\alpha}$, then the radial dimension of the cavity $R_c$
and the average cavity pressure $p_c$ follow 
\begin{equation}
R_c\propto t^{2/(4-\alpha)},~p_c\propto t^{-4/(4-\alpha)}.
\end{equation}
Thus the aspect ratio of the cavity is
$A_c\propto t^{(2-\alpha)/(4-\alpha)}$.
The only free parameter of the extended model is the value of $\alpha$,
and a power-law fit of the ambient density gives a value of
$\alpha = 1.2$ for $r>r_0$.

\begin{figure}
\includegraphics[width=\columnwidth]{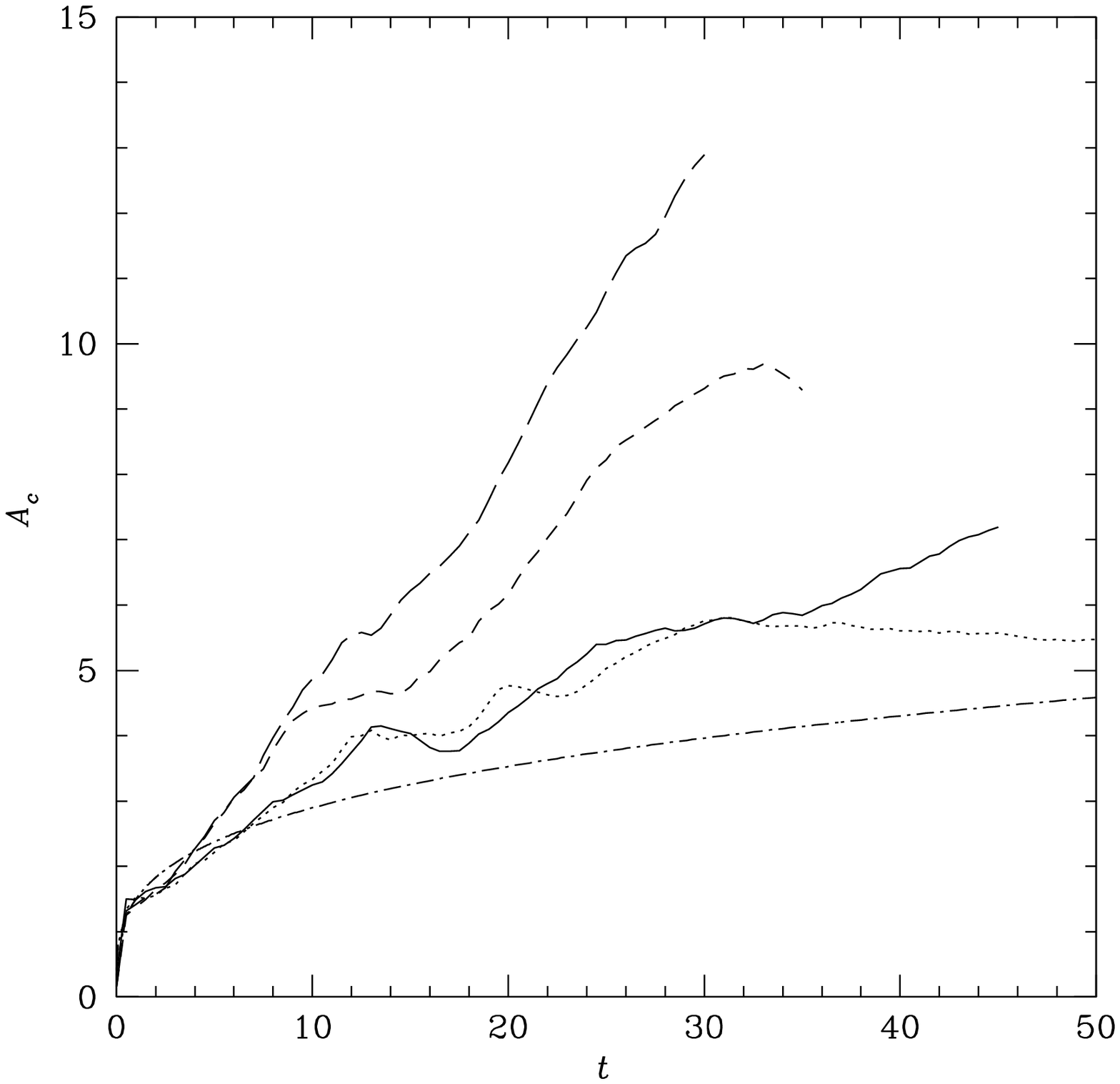}
\caption{The aspect ratio of the cavity as a function of time for all
models.
The solid, dotted, short dashed, and long dashed lines correspond to the
hot leptonic (HL), hot baryonic (HB), cold leptonic (CL), and cold
baryonic (CB) models, respectively.
The dot-short dashed line shows the simple model expectation, where $A_c\propto t^{2/7}$.}
\label{fig5}
\end{figure}

Figure \ref{fig5} shows the evolution of the aspect ratio of the cavity
as a function of time for all models.
The axial length of the cavity $l_c$ increases with time monotonically
until $l_c$ nearly reaches the core radius of the ICM ($l_c\la r_0$).
During this phase the jets propagate with constant velocities of
$v_h\sim0.06$ for hot models and $v_h\sim0.08$ for cold models.
The next phase starts when $l_c$ is greater than the core radius of the
ICM ($l_c>r_0$), which corresponds to $t>7$ for the hot models and $t>5$
for the cold models.
In this phase the jets accelerate progressively and the advance velocity
increases by factors of up to $1.2$ for the hot models and up to $1.6$
for the cold models by the ends of the simulations.
The evolutionary phases can be distinguished in the time evolution of
the aspect ratio of the cavity $A_c$.
During the first phase $A_c$ increases with the predicted evolution
until $t\sim7$ for the hot models and $t\sim5$ for the cold models.
After that, neither the hot models nor the cold models follow the
predicted evolution, mostly because of dynamical processes such as
vortex shedding and mixing that affect $R_c$.
In particular, there are no significant evolutionary differences in
$A_c$ between the leptonic and baryonic models although they have very
different plasma compositions.
 
\begin{figure}
\includegraphics[width=\columnwidth]{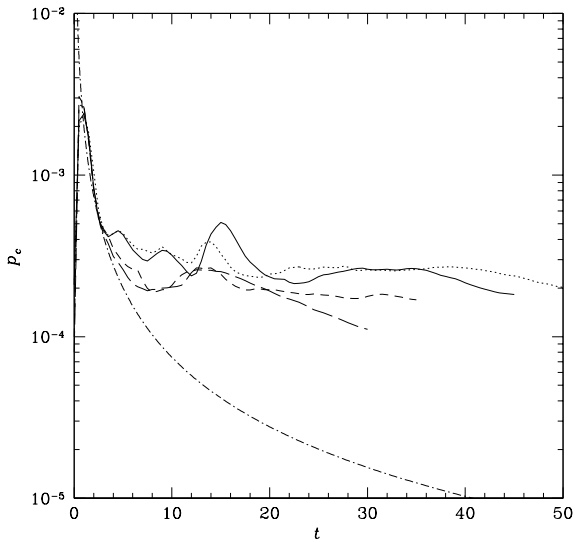}
\caption{Evolutions with time of the average cavity pressure for all
models.
The results for the hot leptonic (HL), hot baryonic (HB), cold leptonic
(CL), and cold baryonic (CB) models are shown by the solid, dotted,
short dashed, and long dashed lines, respectively.
The dot-short dashed line indicates the simple model, where $p_c\propto t^{-10/7}$.}
\label{fig6}
\end{figure}

Figure \ref{fig6} displays the temporal evolutions of the average cavity
pressure of all models together with power-law evolution.
The average cavity pressure $p_c$ decreases as a function of time with a
power-law.
The temporal evolution of $p_c$ is not consistent with that predicted by
the extended model, and the average cavity pressure becomes nearly
constant at the end of these simulations.
Despite the continuous decreases in the cavity pressure, the cavity is
still overpressured with respect to the surrounding medium at the end of
simulations.
Differing jet compositions do not lead to significant differences
between the leptonic and baryonic models in the time evolutions of the
average pressure of the cavity.

\subsection{Radio and X-ray emissions}

Under the assumptions that the AGN jet is optically thin and only the
jet material produces nonthermal radio emission, we can compute the
synchrotron emission in extragalactic jets \citep[e.g.,][]{mio97}.
The synchrotron emissivity at frequency $\nu$ is approximately described
in terms of the local thermal pressure
\begin{equation}
j_\nu\propto p^{(\sigma+3)/2}\nu^{-\sigma},
\end{equation}
where $\sigma$ is the spectral index.
Since a power-law fit with $\sigma = 0.7-0.8$ is very close to steep
spectrum observed in many radio galaxies over large ranges of frequency,
we have used $\sigma = 0.75$ in this work \citep{wil13}.
By integrating the synchrotron emissivity in the volume element
$dV = dxdydz$, the total synchrotron intensity $I_R$ is computed as
$dI_R = \delta^2j_\nu dV$, where $\delta = [\Gamma(1-v\cos\theta)]^{-1}$
is the Doppler boosting factor for a viewing angle $\theta$.

The X-ray emission is entirely dominated by the hot ICM
\citep[e.g.,][]{men11}.
Thermal bremsstrahlung or free-free emissivity is approximated as
\begin{equation}
j_{ff}\propto\rho^2T^{-1/2},
\end{equation}
where $T\propto p/\rho$ is the average temperature of the gas.
The total X-ray intensity $I_X$ is then calculated as $dI_X = j_{ff}dV$
by numerically integrating the free-free emissivity in the volume
element.

\begin{figure}
\includegraphics[width=\columnwidth]{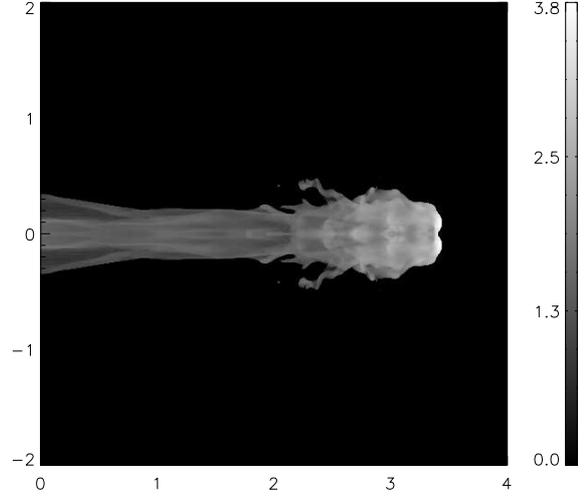}
\caption{The synthetic radio image of the hot leptonic jet (model HL)
for a viewing angle of $90\degr$ at $t = 45$.
The radio intensity map is symmetric with respect to the jet axis and is
shown in arbitrary logarithmic units.}
\label{fig7}
\end{figure}

Figure \ref{fig7} shows the synthetic radio image of the hot leptonic
jet for a viewing angle of $90\degr$ at $t = 45$.
Doppler boosting has a little effect on the emission of the jet at the
viewing angle of $90\degr$, so that the synthetic image is closely
related to the intrinsic emissivity in the jet.
Reducing the viewing angle, excessive Doppler boosting of the beam would
outshine completely the diffuse emission from the cocoon.
The synchrotron emission is dominated by the bright hot spot, where the
pressure is maximum, while the beam and the cocoon show relatively weak
emission features.
The jet itself inflates a radio lobe aligned with the jet axis since
most jet material is around the jet axis.
In all the models, the radio image appears morphologically similar to
observations of the extended radio emission from sources with collimated
radio jets.

\begin{figure}
\includegraphics[width=\columnwidth]{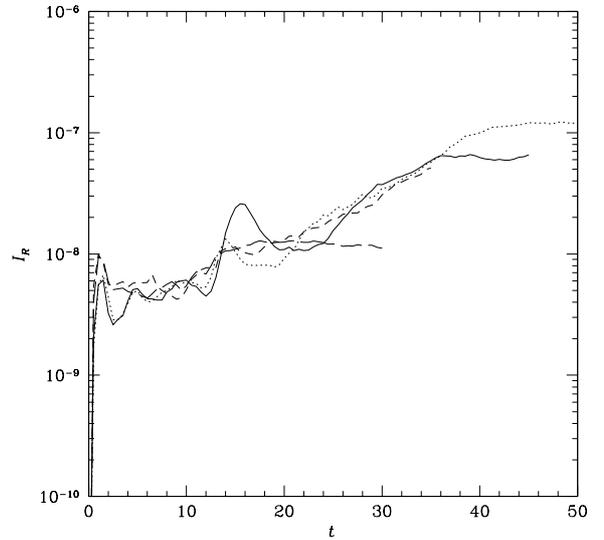}
\caption{The total intensity curves of the radio emissions for all
models.
The hot leptonic (HL), hot baryonic (HB), cold leptonic (CL), and cold
baryonic (CB) models are drawn as the solid, dotted, short dashed, and
long dashed lines, respectively.
The total radio intensities are given in arbitrary units.}
\label{fig8}
\end{figure}

The total intensity curves of the radio emissions for all models are
shown in Figure \ref{fig8}.
Although there are quantitative differences among the models, the total
intensity curves show qualitatively similar trends.
The total radio intensity steeply increases as the jets emerge and build
up the lobes in the nearly constant density medium and then gradually
increases as the jets propagates through the ambient medium with
declining density.
Low-amplitude fluctuations of total radio intensity curves are seen in
late epochs as a result of the onset of shocks.
In our models, the effect of the composition of the jets does not lead
to significant differences in the total intensity curves of the radio
emissions even if the jets have different plasma compositions.

\begin{figure}
\includegraphics[width=\columnwidth]{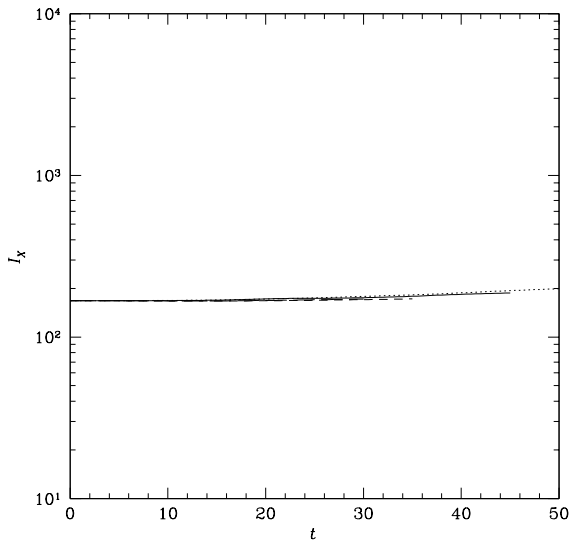}
\caption{The total intensity curves of the X-ray emissions for all
models.
The solid, dotted, short dashed, and long dashed lines correspond to the
hot leptonic (HL), hot baryonic (HB), cold leptonic (CL), and cold
baryonic (CB) models, respectively.
The total X-ray intensities are given in arbitrary units.}
\label{fig9}
\end{figure}

The bright radio lobe in the radio image is visible as dark region in
the X-ray image, which appears morphologically similar to observations
of X-ray cavities \citep[e.g.,][]{ran11}.
This supports the hypothesis that the bow shock surrounding the head of the jet is the
important mechanism for producing X-ray cavities in the ICM.
The dominance of the X-ray emission from the bow shock is evident from
the X-ray emission map as the bow shock outshines the cavity \citep{san16}.
The X-ray emission of the bow shock decreases as the bow shock decreases
in strength.
Figure \ref{fig9} shows the total intensity curves of the X-ray emissions
for each of the models.
The X-ray intensity curves do not fall off for any of the models although the
jets produce the X-ray cavity in the external medium.
This implies that the total X-ray emissions may subside only very gradually after the
jets interact with the ICM.

\section{Discussion and conclusion}

We have performed the first three-dimensional relativistic hydrodynamic
simulations of extragalactic jets with different plasma compositions in
a hydrostatic ICM environment.
The numerical simulations use a general equation of state for a
milti-component relativistic gas that closely approximates the Synge
equation of state for a relativistic perfect gas.
We have considered four models of hot and cold jets made of pure
leptonic and baryonic plasma compositions, propagating in a hydrostatic
intracluster atmosphere with decreasing density and pressure profiles.

In our investigation, morphological and dynamical differences between
the hot and cold jets are much more evident than those between the
leptonic and baryonic jets.
The cold jets propagate at faster velocities than do the hot ones, producing narrower bow shocks
and thiner cocoons. The hot jets with slower advance velocities are
dominated by broader bow shocks and thicker cocoons.
Between the hot models, the leptonic jets advance slightly faster and generate
bow shocks with relatively narrower conical shapes than do the baryonic
jets, but these features are reversed in the cold models.
In spite of very different plasma compositions, the differences in
morphology and dynamics are not significant between the leptonic and
baryonic jets.
In all these models the jets propagate with constant velocities until
the axial length of the cavity nearly reaches the core radius of the ICM
and then the jets accelerate until the jet advance velocities increase
by factors of about $1.2$ to $1.6$ by the end of simulations.
The temporal evolutions of the average cavity pressure does not agree
very well with the power-law prediction of the extended theoretical
model of \citet{sch02} and \citet{per11}.
Differences in the composition of the jet do not lead to significant
differences in the time evolutions of the average pressure of the cavity.

The radio image appears morphologically similar to observations of the
extended radio emission with collimated radio lobes.
The bright radio lobe should appear as a dark region in the X-ray image,
which indicates that the bow shock surrounding the head of the jet is
the important mechanism for the production of X-ray cavities in the ICM.
Although there are quantitative differences among the models the total
radio and X-ray intensity curves show qualitatively similar trends for
all the models we considered.
The total radio intensity steeply increases at early times in the
evolution of the jets as the jets emerge through the ambient medium, but
it then gradually increases with overlaid low-amplitude fluctuations as
the jets propagates through the ICM.
The X-ray intensity curves do not fall off, which implies that the X-ray
emissions may decline only very gradually after the jets produce the X-ray cavity
in the ICM, as compression and heating of the remaining ICM can compensate
for the loss of flux from the lobe region.

We only use one, albeit typical, hydrostatic cluster setup for our
simulations; however, we recognize that the evolution of AGN jets can be
quite different in a dynamic ICM, such as that produced during and
following the merger of clusters \citep{roe97} or a cosmologically
evolved galaxy cluster \citep{mor10}.
Furthermore, the simulations presented here do not include magnetic
fields.
Even if the jet evolution is not dominated by magnetic fields, the
presence of weak magnetic fields should affect the interaction of jet
and cluster material and can change their evolution \citep[e.g.,][]{mck09}.
The full three-dimensional relativistic magnetohydrodynamic simulations
of jets in realistic clusters should be performed in order to
understand the effect of magnetic fields on jet and cluster evolutions
as well as the role of irregular ambient media in the structure and
dynamics of the jets.
Such simulations could be crucial for more complete picture of the
evolution of AGN jets in the ICM and their impact on the galaxy cluster.

\section*{Acknowledgments}

We thank Paul Wiita for useful comments that improved the presentation
of the paper.
This work was supported by the National Research Foundation of Korea
grant funded by the Korean Government (NRF-2009-351-C00029).


\bsp
\label{lastpage}

\end{document}